# Growth of InN quantum dots to nanorods: A competition between nucleation and growth rates

Kishore K. Madapu,[1,*] Sandip Dhara,[1,*] S. Polaki,[1] S. Amirthapandian,[2] and A. K. Tyagi[1]

[1]Surface and Nanoscience Division, Indira Gandhi Centre for Atomic Research, Kalpakkam–603 102, India

[2]Materials Physics Division, Indira Gandhi Centre for Atomic Research, Kalpakkam-603 102, India

**Abstract :** Growth evolution of InN nanostructures via a chemical vapor deposition technique is reported using $In_2O_3$ as precursor material and $NH_3$ as reactive gas in the temperature range of 550–700 $^o$C. Morphology of the nanostructures solely depends on the growth temperature, evolving from quantum dot sized nanoparticles to nanorods. It is found that 630 $^o$C is the threshold temperature for nanorod growth. At 630 $^o$C, nucleation starts with multifaceted particle having {10–12} surface planes. Subsequently, hexagonal polyhedral NRs are grown along the [0001] direction with non–polar surfaces of *m*–planes {10–10}. A comprehensive study is carried out to understand the evolution of nanorods as a function of growth parameters like temperature, time and gas flow rate. Change in the morphology of nanostructures is explained based on the nucleation rate and the growth rates during the phase formation. Raman studies of these nanostructures show that a biaxial strain is developed because of unintentional impurity doping with the increase in growth temperature.

Corresponding Author : Tel.: +91-44-27480015

E-mail address: madupu@igcar.gov.in (Kishore K. Madapu), dhara@igcar.gov.in (Sandip Dhara)

**Introduction**

InN, belonging to the III−V nitride semiconductors, has the superior electronic properties like high mobility and high drift velocity, as compared to the other semiconductor materials because of its low effective electron mass [1]. In addition to this, InN has shown the overshoot in the saturation velocity [2]. Because of this reason InN is considered as a future high−speed electronic device material. InN is reported to have high electron accumulation at the surface region [3]. This unique property can be utilized in the gas sensing applications. In recent years, especially, InN nanostructures are attracted a great deal of attention because of its high surface to volume ratio blended with the property of high surface electron accumulation [4,5]. In addition to these applications, alloying of InN with the other III−V nitride semiconductors can be used to tune the band gap from near infrared to deep UV region (0.7−6 eV). The band gap tuning open the possibility of making the emitters and detectors in a large energy spectrum using a single material with different compositions [6]. Despite its attracting properties, however it is less studied material as compared to other nitrides semiconductors such as GaN and AlN. The reason for InN being a less studied material is because of the difficulty in the synthesis owing to its inherent properties like low dissociation temperature and high equilibrium pressure of N over InN [7,8].

III-V nitrides have been suffered with huge dislocation density in thin films because of the lack of native substrates for these materials [9]. As a result, crystalline quality of these materials reduces which renders the degradation in optical emission properties. Like other III-V nitrides, InN also suffers with lack of native substrates. However, nano−columns of these materials are almost free from dislocations because of their high aspect ratio and efficient strain relaxation through nanoscale lateral dimension [10]. However, nano−columns are still suffering from the threading dislocation near the interface due to the lattice mismatch with substrates. A lot of research is focussed in reducing the dislocation density for these materials to enhance the optical quality.

Even though InN has difficulties in synthesis, reports are available on the growth of InN nanorods (NRs) or nanowires (NWs) with different techniques like molecular beam epitaxy (MBE) [4,5], guided−stream thermal chemical vapor deposition (CVD) [11], controlled−carbonitridation reaction route [12], and thermal CVD [13-15]. Among these techniques CVD is a commercially viable technique for controlled and large scale synthesis. In CVD technique, growth of one dimensional nanostructures are reported following either catalysts assisted vapor−liquid−solid (VLS) mechanism or vapor−solid (VS) mechanism without using catalyst [16]. In VLS mechanism, usually, the growth of NWs is preferred to carry at high temperature because of low vapor pressure of source material at low temperatures. Synthesis of InN via VLS method is limited by the growth temperature because of its thermal instability at high temperature both towards its oxidation to $In_2O_3$ phase [7], and its own compositional integrity [7,8]. InN converts to $In_2O_3$ at the high temperatures in air atmosphere, above 550 $^o$C [7]. Moreover, catalyst assisted VLS growth is difficult with In metal as the source material because of its negligible vapour pressure at ∼ 500 $^o$C. Metal organic precursors are needed for the catalyst assisted growth of InN NWs at low temperatures. However handling of the metal organic precursor is tricky. Secondly, InN phase can be formed in large−scale via VS mechanism with the nitridation of $In_2O_3$ [17-20].

Previously, it was reported that the morphology of InN nanostructures grown via VS mechanism in CVD technique strongly depends on the growth temperature [15,17]. Detailed role of growth parameters of temperature, time and gas flow rate on the morphology of InN nanostructure is hardly reported particularly in the CVD technique. For realization of the device grade material, however, one has to clearly understand the variation in morphology and growth direction of NRs with temperature. Thus a detail understanding of the growth mechanism involving crystallographic analysis is necessary with respect to growth parameters.

In this study, we report growth and evolution of large amount of InN nanostructures in different morphologies via a nearly homogeneous nucleation growth mechanism without using any substrate for the dislocation free growth. Extensive study was carried out with variation of growth parameters like temperature, time and gas flow rate in close steps to understand the growth mechanism of NRs. Crystallographic structural analysis is adopted, for the first time, to establish a growth mechanism for the temperature dependent evolution of morphology in these nanostructures from quantum dots (QDs) to one dimensional NRs grown in the atmospheric CVD technique.

**Experimental**

Growth of InN nanostructures was performed by the nitridation of $In_2O_3$ powder (Purity 99.999%, Sigma Aldrich) with reactive ultra high pure $NH_3$ (5N) at different temperatures. Growth was carried in the customized CVD setup with a horizontal tube furnace via a catalyst free VS process [17-20]. Initially, 40 mg of $In_2O_3$ powder was kept in the ceramic ($Al_2O_3$) boat. Subsequently this boat was transferred to a 1 inch quartz tube and later the tube was degassed by rotary pump to the base pressure of $10^{-3}$ mbar. Temperature of the furnace was increased to growth temperatures at the rate of 20 $^o$C per minute. Nanostructures were grown in the temperature range of 550 to 700 $^o$C. Reactive ultrahigh pure $NH_3$ (99.999%) was introduced at the temperature of 400 $^o$C with constant flow rates. Highly pure Ar was purged with flow rate of 100 sccm until the $NH_3$ introduction. After the reaction, the furnace was cooled to room temperature under the $NH_3$ atmosphere. Black colour powder is collected from the ceramic boat for the characterization.

Field emission scanning electron microscope (FESEM; Zeiss, SUPRA 55) was used to study the morphology of the nanostructures. Detailed structural analysis was performed using X–ray diffraction (XRD, Brucker, D8 Discover) technique and high resolution transmission electron microscopy (HRTEM; Libra 200 Zeiss) along with the selected area electron diffraction (SAED) analysis. TEM assisted elemental analysis was performed using energy dispersive X–ray spectroscopy (EDS) with the aperture size of 20 μm. TEM samples are prepared by dispersing the nanostructures in isopropanol in very dilute amount and subsequently transferring them on amorphous carbon coated copper grid. Vibrational properties were studied by micro–Raman spectrometer (inVia; Renishaw, UK) in the backscattering configuration with the excitation of a 514.5 nm $Ar^+$ laser, 1800 gr/mm grating for monochromatization, and a thermoelectric cooled CCD as detector.

**Results and Discussion**

*Morphological features and structural studies*

Initially, growth of nanostructures was carried out in the temperature range of 550–700 $^o$C at an interval of 50 $^o$C for 4 h each with a flow rate of 100 sccm at an atmospheric pressure. FESEM studies reveal that the morphology of InN nanostructures strongly depends on the growth temperature. Nanostructures grown at 550 $^o$C (Fig. 1a) and 600 $^o$C (Fig. 1b) are found to be multi–faceted nanoparticles. Morphology of these nanostructures is significantly changed to NRs from the nanoparticles as the growth temperature increases to 650 $^o$C (Fig. 1c). The diameter of these NRs is found to be ~100 nm and the lengths up to a several micrometers. Samples grown at 700 $^o$C (Fig. 1d) show presence of micro–tubular and NR structures.

The figure 2 shows the typical TEM images of nanostructures grown at 550 and 650 $^{o}$C. Nanostructures grown at 550 $^{o}$C are composed of InN nanoparticles with typical size less than 8 nm along with larger particles as shown in the FESEM studies (Figs. 2a–2c). Figure 2c shows the high resolution TEM image of nanoparticles in support to the observation of smallest nanoparticles. Exciton Bohr radius of InN is reported to be ~ 8 nm [21]. However, in literature nanoparticles below 30 nm is considered as QDs [22,23]. Lattice spacing of the QD is 0.303 nm, which corresponds to *d* spacing of (10–10) plane of wurtzite InN (inset of Fig. 2c). We also analysed few more QDs having same crystalline planes (supporting information Fig. S1). Nanorods are also shown in the TEM studies (Fig. 2d). As shown in figure 3, growth direction of NRs typically grown at 650 $^{o}$C is established with the help of HRTEM analysis. The variation in the contrast for the TEM images are not correlated to extended defects. The contrasts are due to the variation in thicknesses along the NRs. The FESEM (Figs. 1a, b and inset Fig. 1c) and TEM micrograph (Fig. 3a) reveals that the NRs are bounded by hexagonal surface planes. These surface planes are running parallel to the NR growth direction shown in of figure 3e. At the apex part of the NR, however, surface planes are inclined with an angle of $120^{o}\pm 2^{o}$ (Fig. 3a). HRTEM image of the top plane of the apex part of the NR and its enlarged views are shown in Figs. 3b and 3c, respectively. Lattice spacing of these fringes is found to be 0.304 nm which corresponds to the {10–10} or *m*–planes of InN hexagonal structure. By measuring the angles between the normal of the top and inclined planes of the apex part as shown in figure 3a one can rationalize the crystalline plane of the inclined facets. As depicted in figure 3a, these surface planes are inclined with respect to {10–10} planes. So one can assume that the inclined surface planes can be indexed as {10–1*n*}, where the value of *n* can be derived by measuring the angle between the NR growth direction of [10–10] and the corresponding surface planes. The measured angle between the surface facet plane and {10–10} is found to be $60^{o}\pm 2^{o}$ (Fig. 3a) which is closely matched with the calculated angle between the {10−10} and {10−12}. Thus, it was assumed that nucleation started with [10−10] growth direction and was bounded by {10−12} inclined surface planes. Fast Fourier Transform (FFT) pattern (Figs. 3d) of the corresponding HRTEM images (indicated by red squares in Fig. 3c) shows zone axes along [0001]. Enlarged view of HRTEM image (Fig. 3e) of the middle part of the NR is shown in figure 3f. Lattice spacing between these fringes in the magnified image (Fig. 3g) is found to be 0.285 nm which corresponds to the (0002) planes suggesting the growth of NRs carried along the [0001] or the *c*−axis. As already reported, nano−columns grown along the c-axis is bounded by hexagonal non−polar {10-10} surface planes (*m*-planes) [24]. The results show that the nucleation start with [10−10] direction and growth is carried along the [0001] direction as indicated earlier in Fig. 3d. FFT pattern of different regions, head (Fig. 3d) and middle (Fig. 3h) parts of NR clearly show the different orientation of the planes with zone axes [0001] and [01−10], respectively. Thus the presence of different crystalline planes at different stage of growth reveals the fact that the nucleation and growth process may involve different surface planes.

The XRD study of nanoparticles grown at 550 $^{o}$C (supporting information Fig. S2a) and NRs grown at 650 $^{o}$C (supporting information Fig. S2b) are also used for the confirmation of the crystalline phase. The diffraction peaks corresponding to (10-10), (0002), (10-11), (10−12), (11-10), (10-13), (20-20), (11-12), (20-21), (0004), (20-22) and (10-14) crystalline planes are observed matching with the hexagonal wurtzite phase of InN (JCPDS: 00-050-1239). Nanoparticles grown at 550 $^{o}$C showed the hexagonal wurtzite InN phase along with the cubic $In_2O_3$ phase. The later is observed because of the presence of O (10 atomic %) in the samples grown at 550 $^{o}$C as also recorded in the energy dispersive X–ray spectroscopy (EDS) study (Fig. 4) discussed shortly. In both cases, for the growth temperatures of 550 $^{\rm o}$C and 650 $^{\rm o}$C, we have not observed any substantial changes in the peak positions of diffraction pattern. From the XRD card (JCPDS: 00-050-1239), (10-11) plane is the high intensity reflection of the hexagonal InN phase. We compared the intensity (*I*) ratio of the peaks *I*(10−10)/*I*(10−11) and *I*(0002)/*I*(10−11) in both cases. In case of nanoparticles these ratios are observed with values of 0.384 and 0.375, respectively, which are close enough as reported in the InN XRD card (JCPDS: 00-050-1239). However, in case of NRs we observed the increase in the *I*(10−10)/*I*(10−11) ratio for a value of 0.57364 as compared to the *I*(0002)/*I*(10−11) ratio of 0.17411. Five times increase in the intensity of *m*-plane (10−10) elucidate the fact that NRs growth is along the *c*-axis which is bounded by six polyhedral surfaces of the *m*-planes. Thus, intensity of *m*-plane is increased because of the fact that the NRs, grown along *c*-axis, are lying along the substrate where *m*-planes with six polyhedral surfaces are exposed to X-ray beam. Here *m*-planes are acting as textured kind of orientation with respect single plane normal to *c*-axis. So it is obvious that the XRD intensity will be high for the *m*-planes. Thus, increase in the *m*-plane reflection intensity indicates the growth direction along the *c*-axis which is also predicted from the TEM study as well.

The EDS study was performed along with the HRTEM studies. The EDS analysis shows the presence of off−stoichiometry in InN nanostructures grown at 550 and 700 $^{\rm o}$C (Fig. 4). Presence of oxygen was noticeable in nanostructures grown at low and high temperatures such as 550 $^{\rm o}$C (supplementary Fig. S3) and 700 $^{\rm o}$C. Nanostructures grown at intermediate temperatures, 600 and 650 $^{\rm o}$C, show no deviation from ideal stoichiometry of InN. This may be possible due to the insufficient conversion of $In_2O_3$ to InN at low temperature (550 $^{\rm o}$C). The presence of O at 700 $^{\rm o}$C may be correlated to its associated dissociation of InN and re-oxidation at high growth temperature [7,8]. In the intermediate temperatures, negligible amount of O (1.26 and 2.95 % for 600 and 650 $^{\rm o}$C, respectively) is observed. Thus, in the temperature range of 550 to 650 °C, In to N ratio gradually attains the ideal value of ~1 in InN.

Raman spectral acquisition is carried out at room temperature with micro-Raman spectrometer using 100× microscopic objective of numerical aperture 0.9. Wurtzite phase of InN belongs to the space group of $C_{6v}^4 - P6_3mc$, which have six Raman active modes at the Brillouin zone centre, represented by $\Gamma=2A_1+2E_1+2E_2$ [25,26]. Among these, $A_1$ and $E_1$ modes are polar in nature; because of this reason these modes are further split into longitudinal optic (LO) and transverse optic (TO) components. Non−polar $E_2$ mode, however, is a measure of the biaxial strain in the system [27]. Unpolorized Raman spectra (Fig. 5) of nanostructures showed three prominent symmetry allowed Raman modes in the range of 83–88, 485–489 and 581–593 cm$^{-1}$ which are assigned as the $E_2$(low), $E_2$(high) and $A_1$(LO) modes, respectively [28]. Because of the random orientation of nanostructures with respect to the incident laser one can expect the all possible Raman active modes. We can also observe a tiny peak at the shoulder of $E_2$(high) mode in the range of 471-477 cm$^{-1}$ which can be assigned as the $E_1$(TO) (Fig. 5, marked by ∗). However, the peak is not very clearly observed for the case of 600 °C grown samples. This may be because of the fact that the TO mode is observed in the presence of crystalline defects with increasing deformation potential [29]. The EDS analysis showed the least O content in the sample grown at 600 °C (Fig. 4). Frequencies of $A_1$(LO) and $E_1$(LO) modes fall in the same range of 580-596 cm$^{-1}$. Here we cannot distinguish these peaks because of their occurrence at similar frequencies. However, we assigned the peak around 590 cm$^{-1}$ as the $A_1$(LO) mode because of its asymmetric broadening. Asymmetric broadening is observed in the $A_1$(LO) phonon mode for all these nanostructures. This is because of the coupling of LO−phonons with plasma oscillations of free charge carriers of InN owing to its high electron carrier density [30]. The $E_2$(high) mode is broadened as the growth temperature of nanostructures increases (Fig. 5c). A sharp $E_2$(high) mode with 7 cm$^{-1}$ full width half maxima (FWHM) in the Raman spectrum of nanostructures grown at 550 °C reveals the high crystalinity. Broadening of the $E_2$(high) mode is observed as the growth temperature is increased which revealed the degradation of crystallinity. In addition to these symmetry allowed Raman modes, a peak observed around 441-443 cm$^{-1}$ is assigned as the low frequency plasmon coupled LO phonon mode (L$^-$) [31]. In polar semiconductors, the LO phonons are strongly coupled with the plasmon via microscopic electric field. Coupling of the LO phonons with plasmon leads to the observation of additional modes, such as low frequency and high frequency modes of the LO phonon–plasmon coupled modes (LOPC), namely, $L^-$ and $L^+$, respectively. These frequencies and line shapes are sensitive to the electron density and mobility of carriers in the system [32]. In our study, sharp $L^-$ peaks are observed for all the nanostructures except for the samples grown at 600 °C. This can be correlated with presence of oxygen in the nanostructure grown at 550, 650 and 700 °C. Figure 4 shows the prominent amount oxygen in these nanostructures except 600 °C. Presence of oxygen may contribute the large electron density in system so that the $L^-$ mode acquire phonon like behaviour [33]. Low electron density in nanostructures grown at 600 °C may be the reason for absence of prominent L$^-$ peak. In addition to this, there is a continuous increase in the $A_1$(LO) to $E_2$(high) peak intensity ratio as the growth temperature of nanostructures increases (Fig. 5b). A sharp increase in the $A_1$(LO)/$E_2$(high) is observed from 600 to 650 °C. This increase in the ratio may be attributed to the polarization of LO mode in the $A_1$ symmetry along $c$-axis which is favoured in case a wurtizte unit cell (schematically shown in Fig. 5c) [34]. HRTEM studies also support the fact that these NRs are grown along the $c$-axis at the optimal temperature of 650 °C (Fig. 3).

The FESEM analysis of nanostructures revealed that morphology was significantly changed from multifaceted nanoparticles to NRs when temperature was increased from 600 to 650 °C. Initially growth of nanostructures was carried in the range of 550–700 °C with an interval of 50 °C. In order to study the evolution of NRs from nanoparticles, we carried out growth of nanostructures in the range of 610–640 °C with an interval of 10 °C keeping other growth parameters of gas flow rate and duration of growth as constant. Figure 6 shows the morphology of nanostructures grown in the temperature range of 610–640 °C. Morphology of nanostructures grown at 610 °C shows the continuation of multi–faceted nanoparticles nature (Fig. 6a). At growth temperature of 620 °C, it was observed that agglomeration of nanoparticles was initiated (Fig. 6b). This agglomeration is headed by one of the multifaceted particle as indicated in the figure. However, the morphology of these structures shows corrugated surface features. Nanostructures with NR features were observed at 630 °C with faceted particle at the head (Fig. 6c) indicating a threshold temperature of the NR growth. As the growth temperature increases further to 640 °C the aspect ratio of NRs are observed to increase (Fig. 6d). In order to understand the growth mechanism of NRs, we varied the growth time at optimized growth temperature of 650 °C. It was observed that the growth of NRs is not even started when the growth time is 1 h (Fig. 7a). However, with 2 h growth time NRs are observed with small lengths (Fig. 7b). As the growth time is increased to 3 h (Fig. 7c) and 4 h (Fig. 7d) lengths of NRs are observed to increase. Complete conversion of $In_2O_3$ to InN takes place within the 4 h. Finally, we studied the effect of flow rate of the reactant gas, $NH_3$ at constant growth temperature of 640 °C, close to the optimal value, and growth time of 4 h. Growth of nanostructures were carried out with the flow rates of 30, 50, 70, and 90 sccm of $NH_3$ (Figs. 8a-d). Morphology of nanostructures grown with 30 sccm is observed with nanoparticle nature (Fig. 8a). It was observed that conversion of $In_2O_3$ to InN depends on the flow rates of $NH_3$. The EDS analysis showed a large amount of oxygen (55 % atomic percent) in the same sample (Fig. 9). Presence of O in the nanostructures grown with 30 sccm of $NH_3$ reveals that conversion of $In_2O_3$ to InN is not complete. In the case of nanostructures grown with the flow rates of 50 sccm (Fig. 8b) and 70 sccm (Fig. 7c), however, negligible amount of O (1.61 and 1.03 %, respectively) is observed in the EDS analysis (Fig. 9). Amount of In atomic percentage is slightly increased when flow rate changes from 50 to 70 sccm (Fig. 9). In addition, metallic In was observed at the bottom of ceramic boat when nanostructures were grown with 50 and 70 sccm of $NH_3$ (not shown in figure). In contrary to this, there is no trace of In left in case of NRs grown with 90 sccm (Fig. 8d) and above at 100 sccm as discussed for all previous growth processes (Figs. 1, 6, 7). This observation suggests that, due to the lack of N in the reaction chamber, some of the In is left out without forming the InN when growth is carried out at low flow rates ≤ 70 sccm. Thus 100 sccm of $NH_3$ is optimally accepted value for converting total amount of $In_2O_3$ to stoichiometric InN. Earlier it was reported that morphology of GaN nanostructures grown from $Ga_2O_3$ depends on the $NH_3$ flow rate.[24] In our case, however we failed to observe any significant morphological changes with the flow rate flow rates above 50 sccm (Figs. 8b–8d).

*Growth mechanism of nanorods*

Evolution of nanorod from nanoparticles in the present study is understood in terms of two simultaneous processes of 1) transformation of $In_2O_3$ to InN phase and 2) a temperature dependent nucleation and growth rate. Growth mechanism of InN nanostructures from $In_2O_3$ powder is explained based on the VS mechanism. Earlier reports could not provide an appreciable explanation for the temperature dependent evolution of morphology in InN nanostructures [13,15]. We correlate the evolution of morphology with the nucleation rate and growth rates of newly grown phase during the conversion of $In_2O_3$ to InN at a given temperature. During the phase formation, nucleation and growth are the two important steps. Nucleation ($\dot{N}$) and growth rates ($\dot{G}$) strongly depend on the reaction temperature and diffusion of atomic species which is eventually dependent on the growth temperature. These two rates are competitive when the phase transformation takes place, and domination of one above another will be decided by the transformation temperature. In case of solidification, figure 10a shows the schematic representation of nucleation and growth rate dependence on the temperature. At the lower temperature, the nucleation rate is dominating over the growth rate. In contrary, growth rate dominates at high temperatures. Nucleation rate and growth rates are given below in case solidification from liquid phase [35],

$$\dot{N} = K_1 K_2 K_3 \left[ \exp\left(-\frac{\Delta G^*}{kT}\right) \exp\left(-\frac{Q_d}{kT}\right) \right] \quad \text{...............................} (1)$$

$$\dot{G} = C \exp\left(-\frac{Q}{kT}\right) \quad \text{...............................} (2)$$

where $K_1$ is related to the number of nuclei in solid phase, $K_2$ is the temperature independent constant and $K_3$ is the number of atoms on the nucleus surface in equation (1). In equation (2), $Q$ is the activation energy and $C$ pre–exponential constant which is independent of temperature. These two equations can also be applied to the vapour to solid transformation also. Equation (1) tells that nucleation rate is high at low temperatures with the short range atomic diffusion. Thus diffusion rate at low temperatures limits the growth rates. On the other hand, growth rate is dominated by long range atomic diffusion which takes place at high temperature. At a given reaction temperature, phase transformation depends on both the rates i.e., product of the nucleation rate and the growth rate. All physical phenomena, like grain growth and particle size growth, can be explained based on this competitive nature of these rates. Here, we adopted same mechanism for temperature dependent evolution of morphology of InN nanostructure.

Usually oxide assisted growth (OAG) of NRs or NWs are surrounded by a thin sheath of oxide layer [36]. From the HRTEM analysis we have not observed oxide layer on the NRs.

Moreover, NRs showed abrupt edges (Fig. 3c) of atomic planes which ruled out the OAG of NRs. Usually OAG occurs on the substrate away from the source material [37]. Here, we propose a simple mechanism of phase formation with substitutional chemical reaction at the source material of $In_2O_3$ where O is getting replaced by N. Following possible chemical reactions take place in the reaction chamber. Equation (3) and (4) takes place for the conversion of $In_2O_3$ to InN without any intermediate phase formation [38]. $H_2$ is a present in the chamber as $NH_3$ is cracked above 450 $^o$C [39]. Chemical reactions according to equation (5) takes place at elevated temperature and for long growth time where InN can be dissociated to In droplet and $N_2$.

$$In_2O_3\ (s) + 2H_2(g) \rightarrow In_2O + 2H_2O\ (g) \quad \quad \quad (3)$$

$$In_2O\ (g) + 2NH_3(g) \rightarrow 2InN + H_2O(g) + 2H_2 \quad \quad \quad (4)$$

$$2InN \rightarrow 2In\ (l) + N_2\ (g) \quad \quad \quad (5)$$

Growth temperature plays the key role to decide of the morphology of nanostructures. Earlier it was reported that no appreciable change was observed in the morphology with growth temperature of GaN nanostructures using $Ga_2O_3$ as precursor [24]. Here, however, we have observed that even 10 $^o$C difference can change the morphology of InN nanostructures. At low temperatures atomic diffusion rates are low to prevent the grain growth. Due to this reason nucleation rate is dominating in low growth temperatures of 550 and 650 $^o$C. As a consequence, there can be increased number of nucleation sites with dominating nucleation rate as discussed in describing the equation 1. In the TEM analysis InN nanoparticles are observed with size even less than 8 nm (Fig. 2). Thus it clearly proves that the phase conversion from $In_2O_3$ to InN at temperatures 550 and 600 $^o$C is dominated by the nucleation rate. Had it been a simple conversion of oxide to nitride phase then we would not have observed nanostructure of InN from the commercial $In_2O_3$ powder as precursor. Nucleation rate and its association of short range atomic diffusion are the reasons for the particle nature of nanostructures with a size distribution grown at 550 and 600 °C.

NRs were observed at the growth temperature of 650 °C. This might be possible because of domination of the growth rate over the nucleation rate at high temperature. With the domination of growth rate, formation of fewer nucleus sites takes place and the nucleated clusters grow to large grain size with increasing time. Usually, growth is associated with a long range atomic diffusion and growth rate depends on the rate of diffusion. Evolution of NRs can be understood with the change in the growth temperature at small interval. As previously mentioned, NR growth is initiated with the agglomeration of small particles at a temperature 630 °C with corrugated morphology (Fig. 6). However, multi–faceted nanoparticles play the crucial role in the growth of NRs. At a given temperature, growth starts heterogeneously at certain points and does not occur simultaneously throughout the material [40]. The energy provided by the temperatures at 620 and 630 °C is just sufficient for the agglomeration of particles under one of the multi–faceted nucleating particle. At a temperature of 640 °C, above the threshold, a long range atomic diffusion is dominated for the growth of one-dimensional NRs. Thus the NR growth is achieved by the agglomeration of particles which are headed by few of the multi–faceted nanoparticles shown in figure 6. From the HRTEM analysis we found that the multi–faceted nanoparticles, grown at 550 °C (Fig. 2c) and head region of NRs grown at 650 °C (Fig. 3c), were composed of same (10–10) plane. However, NRs were grown along the direction of [0001] or *c*–axis having a set of polar (0002) plane. Thus, nucleation and growth took place with different planes of the wurtzite InN phase. As shown in the schematic (Fig. 10b), head of the NR is also composed of semi polar {10–12} surface of *r*-planes and the surface contain non-polar {1010} m–planes along the NRs, drawn according to our observations (Fig. 3) and according to Nam *et al* [24]. Diffusion over non–polar planes is solely dependent on the temperature, as there is little chemical affinity of the reacting elements over the saturated InN bond. However, growth will be favoured in the polar planes of the nanostructures. Thus increasing the aspect ratio with increasing growth temperature (Figs. 1 and 7) defines the formation of NRs with growth taking place on the (0001) plane. Similar observation was also made in case of GaN nanostures [41,42]. It was reported that deposition temperature of InN thin film in MBE or vapor phase epitaxy depends on the polarity of the film. The N-polar films can grow at high temperature as compared to the In-polar films [43,44]. These growth techniques involve the reaction of atomic species of In and N on substrate under high vacuum. In the present case, however, InN phase is grown with direct conversation of $In_2O_3$ to InN at atmospheric pressure. Most favourable direction of the reaction is in the right hand side, without any decompositions, until complete conversation takes place because of the presence of large amount of atomic N. Two types of morphology are observed at 700 $^oC$ grown material such as micro-tubes and NRs. Micro-tubes may have formed due to the agglomeration of large number of particles and subsequent growth. Role of oxygen in the 700 $^oC$ grown sample (Fig. 4) may not also be ruled out for the growth of micro-tubular structures.

**Conclusion**

In conclusion we have grown InN nanostructures by nitridation of $In_2O_3$ powder in the catalyst free vapour-solid process. It is observed that the morphology of these nanostructures varies from quantum dots to nanorods depending on the growth temperature. Temperature dependent morphology was explained based on the competitive processes of nucleation rate and growth rates. Quantum dots and nanoparticles, as a nucleation rate dominated process, are grown at relatively low temperatures. Nanorods are grown at high temperature due to long–range atomic diffusion through non–polar surface where growth rate dominates.

Nucleation starts with a multifaceted particles having polar surface planes and growth is carried over non–polar surface planes. Broadening of $E_2$(high) mode in the Raman analysis show development of a biaxial strain in these sample because of unintentional impurity doping with the increase in the growth temperature. Observed enhancement in the peak intensity of polar $A_1$(LO) mode with respect to that for non-polar $E_2$(high) mode with increasing growth temperature show the preferred growth orientation along *c*-axis, as the polarization axis for $A_1$ symmetry in the wurtzite phase. The presumption is also supported by structural analysis which shows growth direction of NRs as *c*-axis.

**Acknowledgements**

We thank Arindam Das of the Surface and Nanoscience Division, IGCAR for fruitful discussion during the manuscript preparation. We thank R. M Sarguna, Condensed Matter Physics Division, IGCAR for her help in XRD studies. We also thank K. Shalini, NIT Trichy, V. T. Antos Shiny, Sarah Tucker College, Tirunelveli and Ahmed Aslam V. V, VIT University, Vellore for their help in experimental work.

**References**

1   A. G. Bhuiyan, A. Hashimoto, A. Yamamotoa, *J. Appl. Phys.* 2003, **94**, 2779.

2   B. E. Foutz, S. K. O'Leary, M. S. Shur, L. F. Eastman, *J. Appl. Phys.* 1999, **85**, 7727.

3   I. Mahboob, T. D. Veal, C. F. McConville, H. Lu, W. J. Schaff, *Phys. Rev. Lett.* 2004, **92**, 036804.

4   N. J. Ku, J. H. Huang, C. H. Wang, H. C. Fang, C. P. Liu, *Nano Lett*. 2012, **12**, 562.

5   T. Stoica, R. J. Meijers, R. Calarco, T. Richter, E. Sutter, H. Luth, *Nano Lett.* 2006, **6**, 1541.

6   T. Kuykendall, P. Ulrich, S. Aloni, P. Yang, *Nat. Mater.* 2007, **6**, 951.

7    Q. Guo, O. Kato, A. Yoshida, *J. Appl. Phys.* 1993, **73**, 7969.

8   B. Onderka, J. Unland, R. S. Fetzera, *J. Mater. Res.* 2002, **17**, 3065.

9   T. Araki, S. Ueta, K. Mizuo, T. Yamaguchi, Y. Saito, Y. Nanishi, *phys. stat. sol. (c)* 2003, **0**, 2798.

10  S. D. Hersee, X. Sun, X. Wang, *Nano Lett*. 2006, **6**, 1808.

11  M. Hu, W. Wang, T. T. Chen, L. Hong, C. Chen, C. Chen, Y. Chen, K. Chen, L. Chen, *Adv. Funct. Mater.* 2006, **16**, 537.

12  L.-W. Yin, Y. Bando, D. Golberg, M.-S Li, *Adv. Mater.* 2004, **16**, 1833.

13  M. C. Johnson, C. J. Lee, E. D. Bourret-Courchesne, S. L. Konsek, S. Aloni, W. Q. Han, A. Zettl, *Appl. Phys. Lett*. 2005, **85**, 5670.

14  Z. H. Lan, W.M. Wang, C. L. Sun, S. C. Shi, C. W. Hsu, T. T. Chen, K. H. Chen, C. C. Chen, Y. F. Chen, L. C. Chen, *J. Cryst. Growth* 2004, **269**, 87.

15  B. Schwenzer, L. Loeffler, R. Seshadri, S. Keller, F. F. Lange, S. P. DenBaarsb, U. K. Mishra, *J. Mater. Chem.* 2004, **14**, 637.

16  N. Wang, Y. Cai, R. Q. Zhang, *Mater. Sci. Eng., R* 2008, **60**, 151.

17      W.S. Jung, C. S. Ra, B. K. Min, *Bull. Korean Chem. Soc*. 2005, **26**, 1354.

18      J. Zhang, L. Zhang, X. Penga, X. Wanga, *J. Mater. Chem*. 2002, **12**, 802.

19      S. Luo, W. Zhou, W. Wang, Z. Zhang, L. Liu, X. Dou, J. Wang, X. Zhao, D. Liu, Y. Gao, L. Song, Y. Xiang, J. Zhou, S. Xie, *Appl. Phys. Lett*. 2005, **87**, 0631091.

20      S. Luo, W. Zhou, Z. Zhang, X. Dou, L. Liu, X. Zhao, D. Liu, L. Song, Y. Xiang, J. Zhou, S. Xie, *Chem. Phys. Lett*. 2005, **411**, 361.

21      P. K. B. Palomaki, E. M. Miller, N. R. Neale, *J. Am. Chem. Soc.* 2013, **135**, 14142.

22      B. Maleyre, O. Briot, S. Ruffenach, *J. Cryst. Growth* 2004, **269**, 15.

23      O. Briot, B. Maleyre, S. Ruffenach, *Appl. Phys. Lett.* 2003, **83**, 2919.

24      C. Y. Nam, D. Tham, J. E. Fischera, *Appl. Phys. Lett.* 2004, **85**, 5676.

25      S. Sahoo, M. S. Hu, C. W. Hsu, C. T. Wu, K. H. Chen, L. C. Chen, A. K. Arora, S. Dhara, *Appl. Phys. Lett.* 2008, **93**, 2331161.

26      J. Segura-Ruiz, N. Garro, A. Cantarero, *Phys. Rev. B* 2009, **79**, 115305.

27      X. Wang, S. Che, Y. Ishitani, A. Yoshikawa, *Appl. Phys. Lett.* 2006, **93**, 171907.

28      K. K. Madapu, N. R. Ku, S. Dhara, C. P. Liu, A. K. Tyagi, *J. Raman Spectrosc.* 2013, **44**, 791.

29      S. Dhara, A. K. Arora, S. Bera, and J. Ghatak *J. Raman Spectros.* 2010, **41**, 1102.

30      T. Kozawa, T. Kachi, H. Kano, Y. Taga, M. Hashimoto, N. Koide, K. Manabe, *J. Appl. Phys.* 1994, **75**, 1098.

31      Y. Cho, M. Ramsteiner, O. Brandt, *Phys. Rev. B* 2012, **85**, 195209.

32      D. Wang, C. -C. Tin, J. R. Williams, M. Park, Y. S. Park, C. M. Park, T. W. Kang, W. -C. Yang, *Appl. Phys.Lett.* 2005, **87**, 242105.

33      R. Cuscó, D. Amador, L. Artús, T. Gotschke, K. Jaganathan, T. Stoica, R. Calarco, *Appl. Phys. Lett*. 2010, **97**, 221906.

34       H. Harima, *J. Phys.: Condens. Matter* **2002,** *14*, R967.

35       W. D. Callister Jr, *Materials Science and Engineering an Introduction*, Seventh Edition,  John Wiley & Sons, Inc., New York, 2007; Chapter 10, pp 311.

36       B. Mandl, J. J. Stangl, E. Hilner, A. A. Zakharov, K. Hillerich, A. W. Dey, L. Samuelson, G. Bauer, K. Deppert, A. Mikkelsen, *Nano Lett.* 2010, **10**, 4443.

37       W. S. Shi, Y. F. Zheng, N. Wang, C. S. Lee, S. T. Lee, *Chem. Phys. Lett*. 2001, **345**, 377.

38       J. Jian, X. L. Chen, M. He, W. J. Wang, X. N. Zhang, F. Shen, *Chem. Phys. Lett*. 2003, **368**, 416.

39       A. H. White, W. Melville, *J. Am. Chem. Soc.* 1905, **27**, 373.

40       J. Ristic, E. Calleja, S. F. Garrido, L. Cerutti, A. Trampert, U. Jahn, K. H. Ploog, *J. Cryst. Growth* 2008, **310**, 4035.

41       P. Sahoo, S. Dhara, S. Amirthapandian, M. Kamruddin, S. Dash, B. K. Panigrahi, A. K. Tyagi, *Cryst. Growth Des*. 2012, **12**, 2375.

42       P. Sahoo, J. Basu,  S. Dhara,  H. C. Fang, C.-P. Liu, T. R. Ravindran, S. Dash, A. K. Tyagi, *J. Mater. Sci.* 2012, **47**, 3447.

43        K. Xu and A. Yoshikawaa, *Appl. Phys. Lett*. 2003, **83**, 251.

44       K. Xu, W. Terashima, T. Hata, N. Hashimoto, Y. Ishitani, A. Yoshikawa, *phys. stat. sol. (c)* 2002, **0,** 377.
20

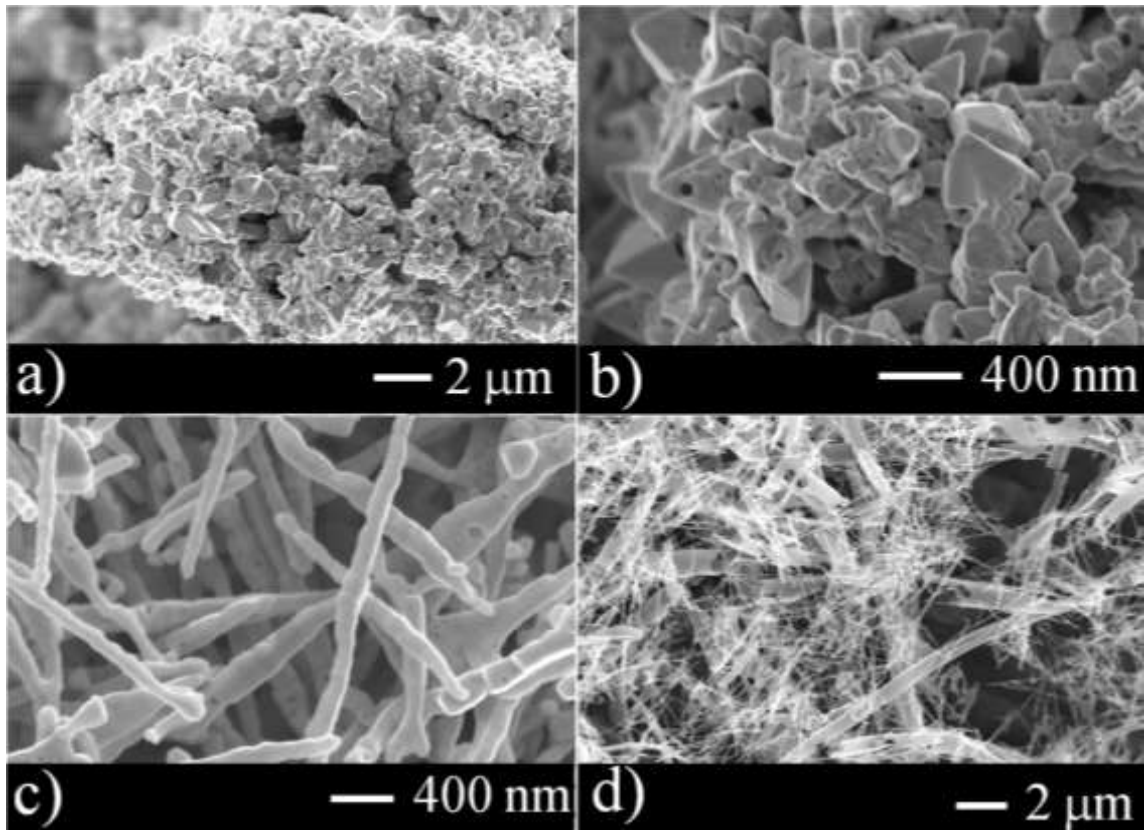

Fig. 1. Morphology of InN nanostructures grown for the temperatures a) 550 °C, b) 600 °C, c) 650 °C, and d) 700 °C.

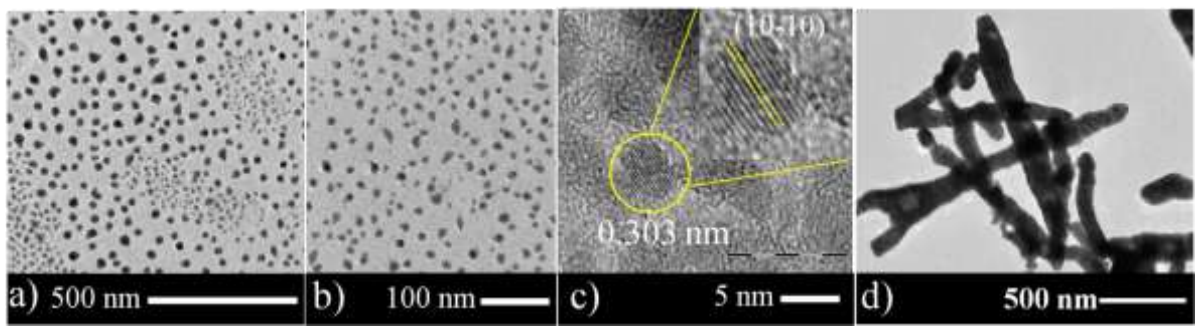

Fig. 2. Low and high resolution TEM images of nanostructures grown at 550 and 650 °C. (a) to (c) are showing the size distribution of nanoparticles grown at 550 °C (d) Low magnification image of NRs grown at 650 °C.

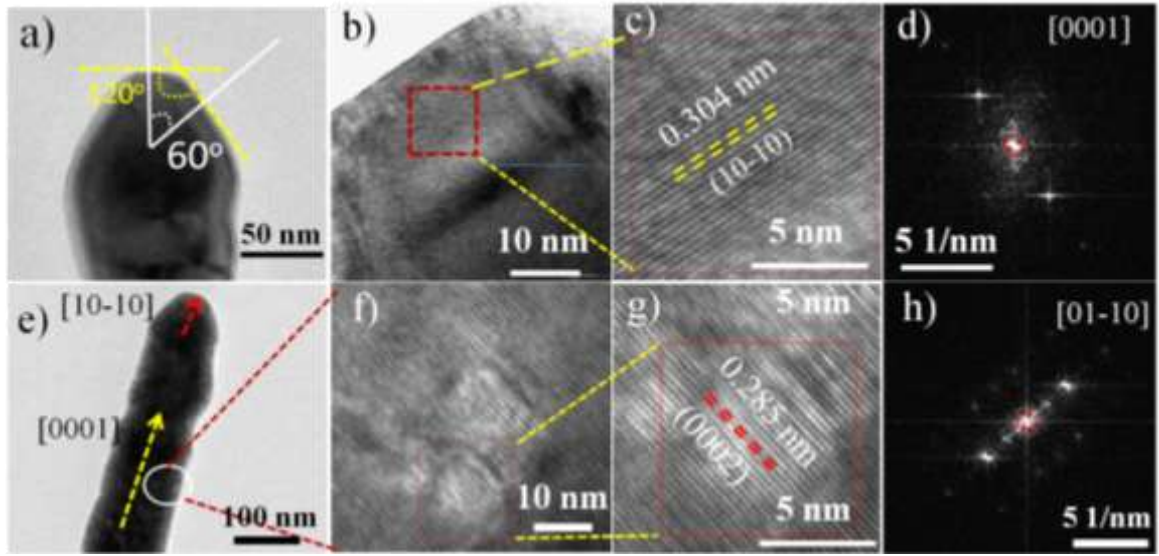

Fig. 3. Detailed HRTEM analysis of NR. a) Low magnification image of apex part of NR which shows the angle between growth direction and surface facet planes is 60°. c) & d) HRTEM images of apex part of NR and its enlarged view, respectively, shows the (10−10) lattice fringes. d) FFT of apex part of the NR (area is indicated by red square in Fig. c). e) Low magnification image of NR shows the growth direction of middle part. f) & g) High resolution image of middle part of the NR and its enlarged view, respectively, shows the lattice fringes of (0002) planes. h) FFT of middle part of the NR (area is indicated by red square in Fig. g). The contrasts in the micrographs are due to the variation in thicknesses along the NRs.

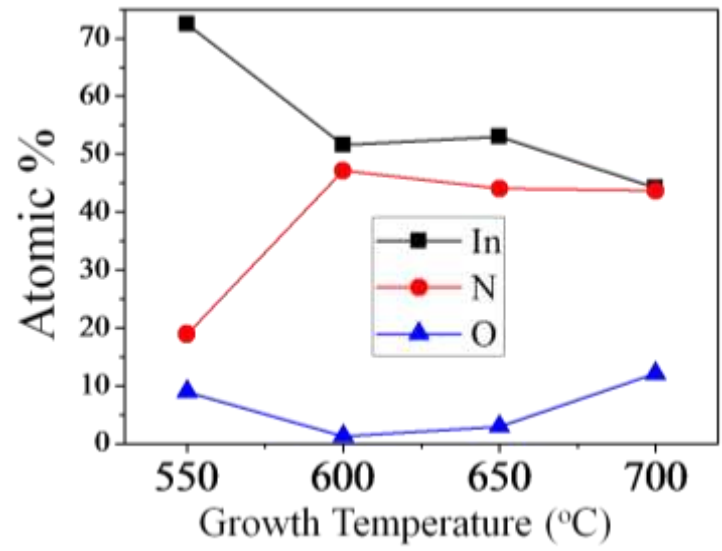

Fig. 4. Variation of atomic percentages of In, N and O in nanostructures grown at different temperatures.

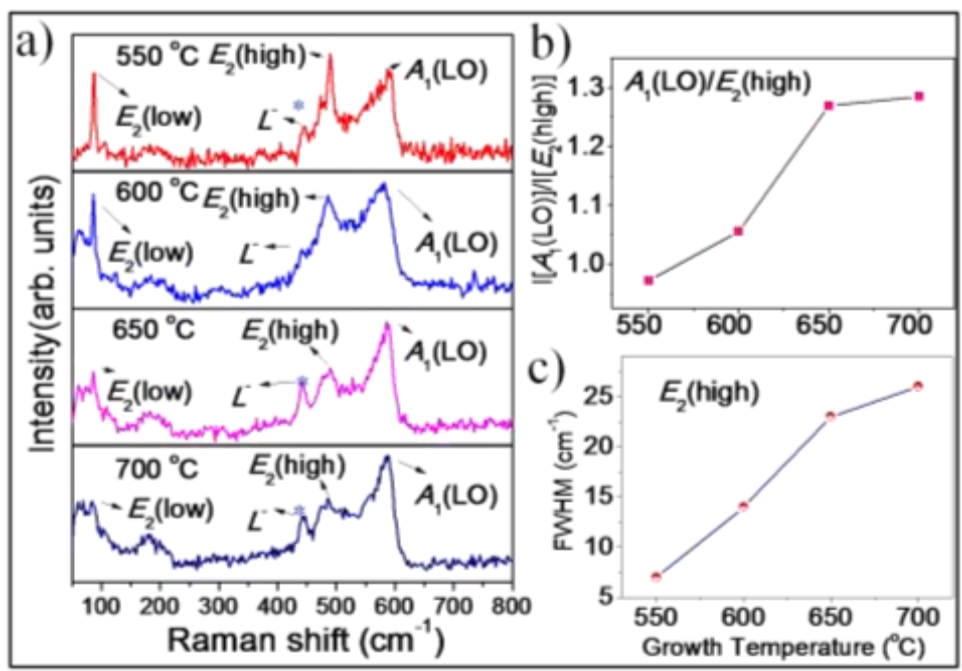

Fig. 5. Raman spectra of nanostructures grown at the temperatures of (a) 550–700 °C. (b) I($A_1$(LO))/ I($E_2$(high)) with different growth temperatures. Lines are drawn as a guide to an eye (c) Variation of $E_2$ mode FWHM with respect to growth temperature. A tiny peak at the shoulder of $E_2$(high) mode in the range of 471-477 cm$^{-1}$ is assigned as the $E_1$(TO) (marked by ∗). $L^-$ is denoted as one of the LOPC modes.

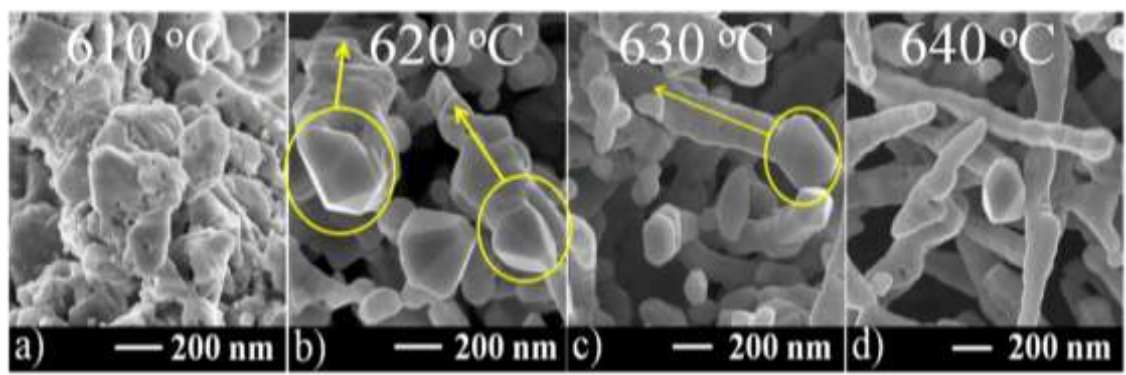

Fig. 6. Evolution of NRs with increase in growth temperatures (a) 610 °C, (b) 620 °C, (c) 630 °C, and (d) 640 °C.

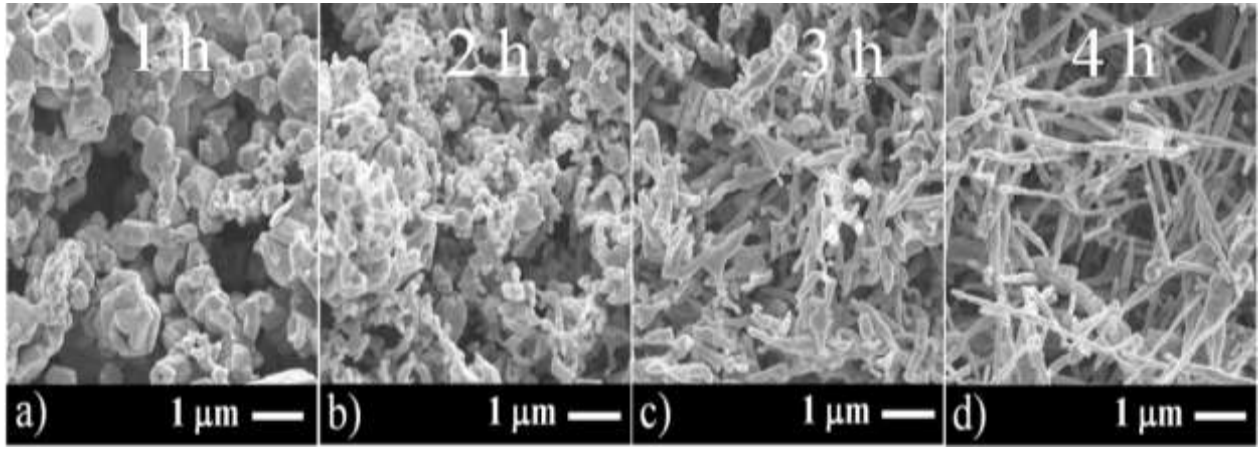

Fig. 7. Morphology variation with growth time of (a) 1h, (b) 2h, (c) 3h, (d) 4h at an optimized growth temperature 650 °C.

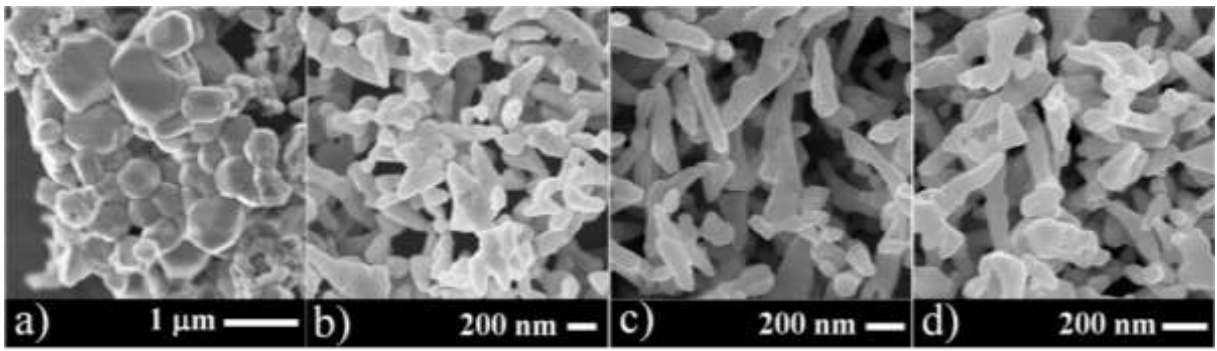

Fig. 8. Optimization of flow rates (a) 30 sccm, (b) 50 sccm, (c) 70 sccm, (d) 90 sccm at a growth temperature of 640 °C, close to the optimal value. No significant change observed in the morphology above 50 sccm.

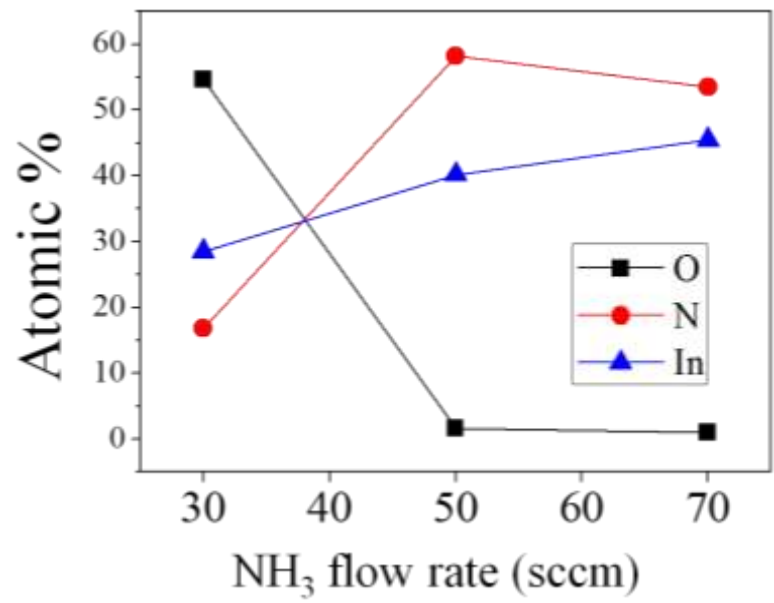

Fig. 9. Variation of atomic percentages of In, N and O in nanostructures grown with different flow rates at 640 °C.

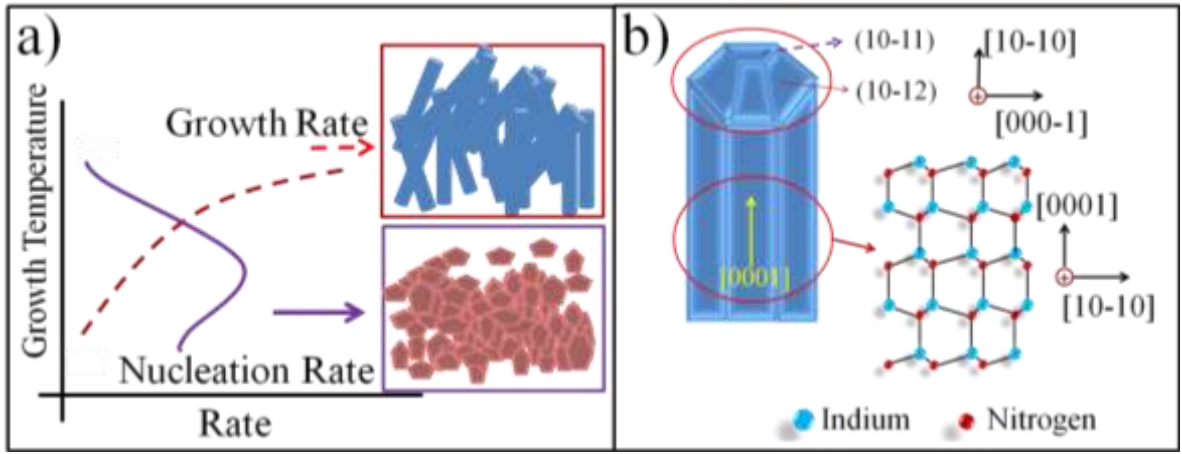

Fig. 10. (a) Graphical illustration of nucleation and growth rates which determines the morphology of nanostructures. (b) Schematic of atomic planes arrangement of NR at different parts. At apex (nucleation part) of NR contains {10−12} semi polar surfaces. These planes can have potential barriers for adatom diffusion because of the polarity. NRs grow along the [0001] direction such that surface have non−polar planes i.e., {1010} at high temperature (650 °C) where growth rate dominates. Diffusion of adatoms over the non-polar planes depends on temperature only. At high temperature, NRs grow with long range atomic diffusion over non-polar surfaces.

Supporting Information :

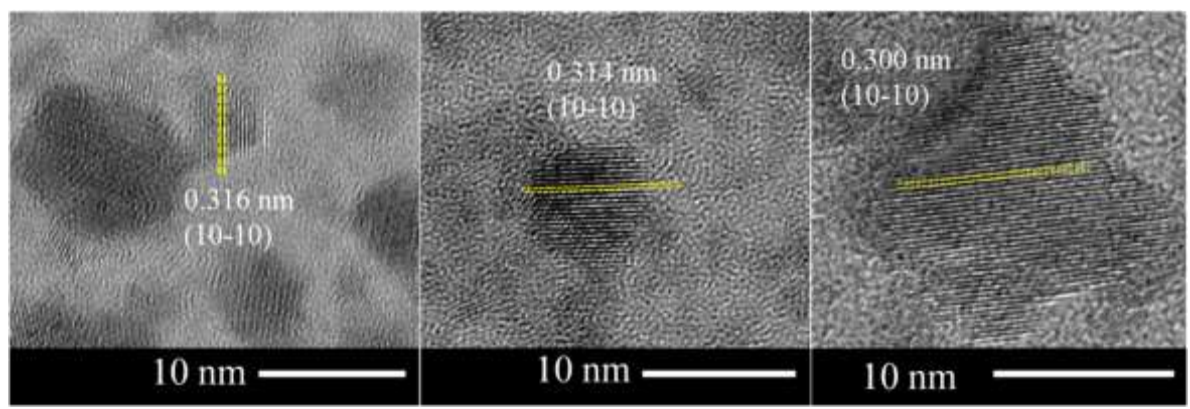

Fig. S1. HRTEM analysis of different quantum dots showing the presence of similar (10–10) planes.

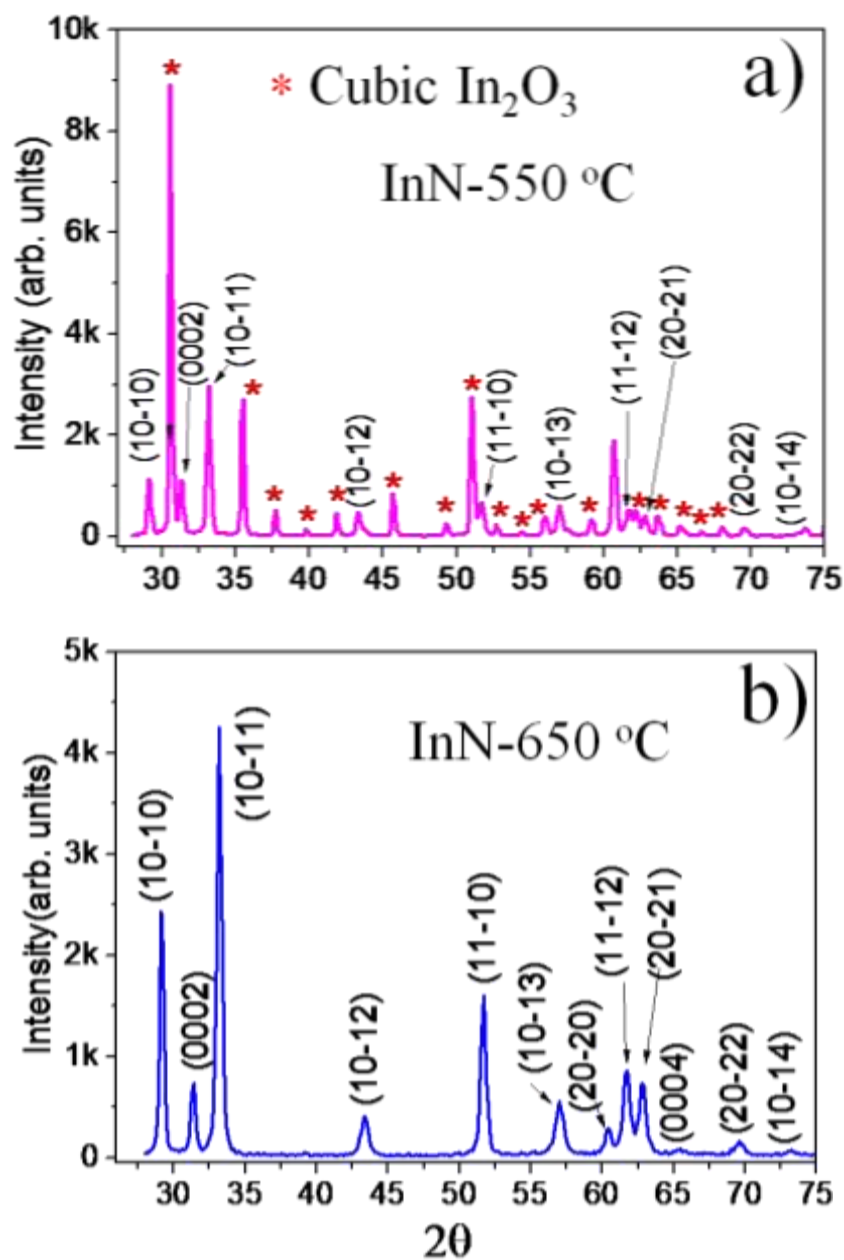

Fig. S2. XRD spectra with diffraction peaks corresponding to (10-10), (0002), (10-11), (10-12), (11-10), (10-13), (20-20), (11-12), (20-21), (0004), (20-22) and (10-14) crystalline planes corresponding to wurtzite InN phase (JCPDS # 50–1239) of typical a) nanoparticles grown at 550 °C along with the substantial presence of cubic $In_2O_3$ and b) NRs grown at 650 °C.

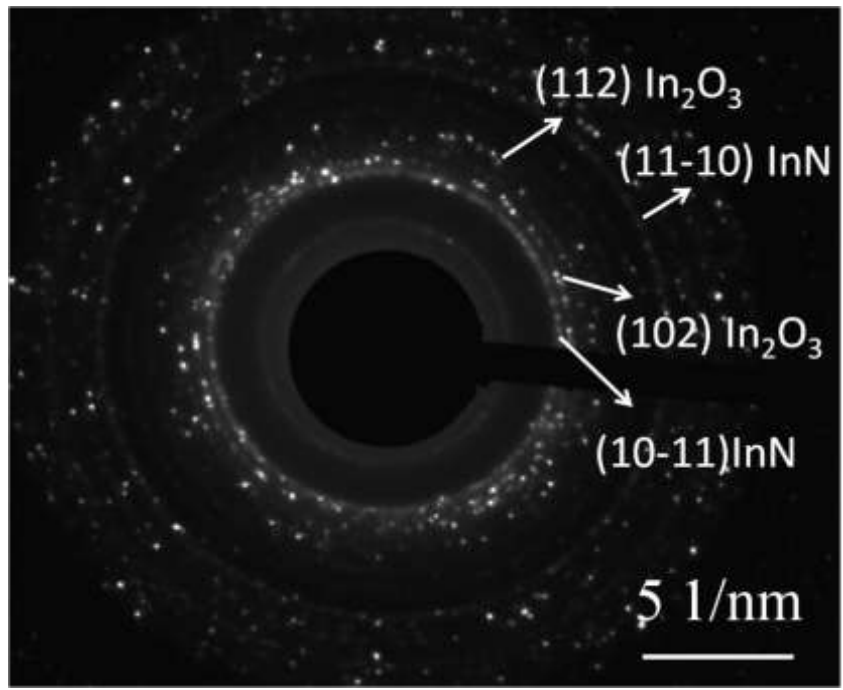

Fig. S3. SAED pattern of nanostructures grown at low temperature of 550 °C showing the crystalline planes corresponding to InN and $In_2O_3$ phases.